\documentclass[traditabstract]{aa}

\usepackage{amssymb, amsmath}
\usepackage[below]{placeins}
\usepackage{graphicx,natbib,txfonts}
\usepackage{verbatim}
\usepackage{subfigure}

\graphicspath {{}}

\begin{document}

\title{Joint reconstruction of galaxy clusters from gravitational lensing and thermal gas}
\subtitle{I. Outline of a non-parametric method}
\author
 {Sara Konrad \and Charles L. Majer \and Sven Meyer \and Eleonora Sarli \and Matthias Bartelmann}
\institute
 {Universit\"at Heidelberg, Zentrum f\"ur Astronomie, Institut f\"ur Theoretische Astrophysik, Philosophenweg 12, 69120 Heidelberg, Germany}

\date{\today}

\abstract{
We present a method to estimate the lensing potential from massive galaxy clusters for given observational X-ray data. The concepts developed and applied in this work can easily be combined with other techniques to infer the lensing potential, e.g. weak gravitational lensing or galaxy kinematics, to obtain an overall best fit model for the lensing potential. After elaborating on the physical details and assumptions the method is based on, we explain how the numerical algorithm itself is implemented with a Richardson-Lucy algorithm as a central part. Our reconstruction method is tested on simulated galaxy clusters with a spherically symmetric NFW density profile filled with gas in hydrostatic equilibrium. We describe in detail how these simulated observational data sets are created and how they need to be fed into our algorithm. We test the robustness of the algorithm against small parameter changes and estimate the quality of the reconstructed lensing potentials. As it turns out we achieve a very high degree of accuracy in reconstructing the lensing potential. The statistical errors remain below $2.0\,\%$ whereas the systematical error does not exceed $1.0\,\%$.}


\maketitle

\section{Introduction}
The core structure of galaxy clusters provides important cosmological information. Based on numerical simulations, we expect the dark-matter distribution to follow a universal profile with characteristic gradients and a scale radius \citep{NFW1997}. Outside relatively small central regions, cluster density profiles should not be strongly affected by baryonic physics because of the long cooling times in the intracluster plasma. Cold dark matter is expected to clump on virtually arbitrarily small scales. The level of substructure in clusters thus potentially constrains the nature of the dark-matter particles \citep[][]{Boylan2009, Gao2011}.

The ratio between the scale and the virial radii of galaxy clusters, dubbed the concentration parameter, is frequently observed to be substantially different than theoretically expected. In particular in strongly gravitationally lensing clusters, concentration parameters significantly higher than those found in numerical simulations have been detected \citep[][Fig.~14]{Broadhurst2008, Coe2012}. It is fundamentally important to find out whether this discrepancy reflects insufficient understanding at the level of our theory of cosmological structure formation, or whether it is a combination of baryonic physics, selection effects and measurement biases that gives rise to this observation.

A multitude of precise observational data on galaxy clusters is or is becoming available: Weak and strong gravitational lensing constrain the distribution of the total matter density projected along the line-of-sight. X-ray emission and the thermal Sunyaev-Zel'dovich effect constrain the density, temperature and pressure of the intracluster gas. Galaxy kinematics constrain the gradient of the gravitational potential, albeit in a fairly entangled way. We should expect to find the strongest constraints on the core structure of galaxy clusters by combining all available types of data in a common and consistent way.

Without equilibrium or symmetry assumptions, only data from gravitational lensing can be interpreted, while the interpretation of gas physics and galaxy kinematics requires at least equilibrium assumptions. Given such assumptions, however, all types of galaxy cluster data can be theoretically modelled on the basis of the gravitational potential. In this first paper of an intended series, we are focussing on X-ray emission, devising an algorithm to convert the observed surface brightness profile into a projected potential that can then directly be combined with data from gravitational lensing. Studies in progress, to be reported on in due course, will concern the thermal Sunyaev-Zel'dovich effect and galaxy kinematics. The ultimate goal of our studies is a non-parametric method combining strong and weak lensing, observations of thermal gas physics and galaxy kinematics into one consistent model for the projected cluster potential.

This paper is structured as follows: In Sect.~2, we develop an algorithm \citep[see][]{Lucy1974, Lucy1994} for reconstructing the projected cluster potential from the X-ray surface brightness. Numerical tests, described in Sect.~3, illustrate how this algorithm performs under reasonably realistic conditions. Although we adopt a spherically symmetric cluster potential for this test, spherical symmetry is not a necessary condition for our algorithm to work. The influence of a possible deviation from spherical symmetry is exemplified in Sect.~3.3. The results and our conclusions are summarised in Sect.~4.

\section{Recovering the projected gravitational potential from X-ray surface brightness}

\subsection{Basic relations}
The central object of our study is the Newtonian gravitational potential $\Phi$. Gravitational lensing measures the projection
\begin{equation}
\label{eq:01}
  \psi(\vec{\theta}) = 2\frac{D_\mathrm{ls}}{D_\mathrm{l}D_\mathrm{s}}\int\mathrm{d} z\,\Phi(D_\mathrm{l}\vec{\theta}, z)\;,
\end{equation} 
geometrically weighted by a combination of the angular-diameter distances $D_\mathrm{l, s, ls}$ from the observer to the lens, to the source, and from the lens to the source, respectively. Different lensing observables characterise derivatives of $\psi$ of different order. Time-delay measurements constrain differences in $\psi$ along different lines-of-sight, deflection-angle differences between components of a multiple image constrain differences in the gradient of $\psi$, elliptical distortions constrain the curvature matrix of $\psi$, and flexion will hopefully soon constrain its third-order derivatives. Combining all available lensing observables thus naturally leads to a reconstruction of the lensing potential $\psi$ which can be more detailed where observables sensitive to higher-order derivatives can be measured.

At least in or near hydrostatic equilibrium, the density and temperature of gas in the lensing potential well are also fully characterised by the Newtonian potential. We begin with the hydrostatic equation
\begin{equation}
\label{eq:02}
  \vec{\nabla} P = -\rho\vec{\nabla}\Phi
\end{equation}
and assume that the gas satisfies the polytropic relation
\begin{equation}
  \frac{P}{P_0} = \left(\frac{\rho}{\rho_0}\right)^\gamma,
\label{eq:03}
\end{equation} 
with an effective adiabatic index $\gamma$. Assuming spherical symmetry, equation (\ref{eq:02}) is immediately integrated to give
\begin{equation}
  \left(\frac{\rho}{\rho_0}\right)^{\gamma-1} = \frac{\gamma-1}{\gamma}\frac{\rho_0}{P_0}(\Phi_\mathrm{cut}-\Phi)\;,
\label{eq:04}
\end{equation}
where quantities with a subscript $0$ refer to an arbitrary radius $r_0$ which could, for example, be set to zero. For practical reasons, we introduce a cutoff radius $r_\mathrm{cut} > r_0$ and fix the potential such that $\Phi_\mathrm{cut} - \Phi(r_\mathrm{cut}) = 0$.

For $\gamma > 1$ and a density profile that decreases monotonically\footnote{It can be seen below that both cases are fulfilled in our consideration.}, $\Phi(r)$ can be arranged to be negative for $r<r_0$ by the structure of Eq. (\ref{eq:04}). Therefore $\rho$ remains positive and semi-definite. The quantity
\begin{equation}
  \gamma\frac{P_0}{\rho_0} = c_\mathrm{s, 0}^2
\label{eq:05}
\end{equation} 
appearing in Eq.~(\ref{eq:04}) is the squared sound speed at the cutoff radius. For convenience, we introduce the dimension-less potential
\begin{equation}
  \varphi = \frac{\gamma-1}{c_\mathrm{s, 0}^2}\left(\Phi_\mathrm{cut}-\Phi\right)
\label{eq:06}
\end{equation} 
and obtain the gas density
\begin{equation}
  \rho = \rho_0\varphi^{1/(\gamma-1)}\;.
\label{eq:07}
\end{equation}
The temperature of an ideal gas in thermal equilibrium with the potential $\varphi$ is
\begin{equation}
  T = \frac{\bar m}{k_\mathrm{B}}\frac{P}{\rho} =
  \frac{\bar m}{k_\mathrm{B}}\frac{P_0}{\rho_0}\varphi =
  T_0\varphi\;,
\label{eq:08}
\end{equation}
where $\bar m$ is the mean gas-particle mass and $k_\mathrm{B}$ is Boltzmann's constant.

Since the frequency-integrated emissivity due to bremsstrahlung is given by
\begin{equation}
  j_X = C\rho^2T^{1/2}\;,
\label{eq:09}
\end{equation}
it can be related to the potential by
\begin{equation}
  j_X = C\rho_0^2T_0^{1/2}\varphi^\eta\;,\quad
  \eta = \frac{3+\gamma}{2(\gamma-1)}\;.
\label{eq:10}
\end{equation}
For realistic effective adiabatic indices $1.1\lesssim\gamma\lesssim1.2$ \citep{Finoguenov2001}, the exponent $\eta$ is quite a large number, $10\lesssim\eta\lesssim20$.

Equation (\ref{eq:10}), together with the fact that ordinary lensing effects are determined by second-order derivatives of the projected Newtonian potential, suggests the following algorithm for combining X-ray and lensing data:

\begin{enumerate}

\item By deprojection of an X-ray surface brightness map $S_X$, find an estimate $\tilde j_X$ for the X-ray emissivity $j_X$. How this could be done e.g.\ by means of Richardson-Lucy deprojection will be discussed below.

\item Use Eq.~(\ref{eq:10}) to infer an estimate
\begin{equation}
  \tilde\varphi = \left(\frac{\tilde j_X}{C\rho_0^2T_0^{1/2}} \right)^{1/\eta}
  \label{eq:11}
\end{equation} 
for the three-dimensional, scaled Newtonian potential.
  
\item Project $\tilde\varphi$ along the line-of-sight to obtain an estimate $\tilde\psi$ for the two-dimensional potential, which is proportional to the lensing potential and can thus directly be combined with estimates of $\psi$ derived from lensing.

\end{enumerate}

Since $\eta$ is large, $1/\eta\ll1$, which is a most welcome property of Eq.~(\ref{eq:11}): Inevitable fluctuations in the deprojected estimate $\tilde j_X$ will be considerably smoothed that way.

\subsection{Deprojection}

Different algorithms exist for the deprojection of two- into three-dimensional distributions. Without symmetry assumptions, such algorithms cannot be unique. Assuming spherical symmetry for simplicity, a three-dimensional function $f(r)$ is related to its two-dimensional projection $g(s)$ by
\begin{equation}
  g(s) = \int_{-\infty}^\infty\mathrm{d} z\,f\left(\sqrt{s^2+z^2}\right)\;,
\label{eq:12}
\end{equation} 
where $s$ is the projected radius and the coordinate system is chosen such that the $z$-axis points along the line-of-sight. At fixed $s$, we have $z\mathrm{d} z = r\mathrm{d} r$, allowing us to transform Eq.~(\ref{eq:12}) to
\begin{equation}
  g(s) = 2\int_s^\infty\frac{r\mathrm{d} r}{\sqrt{r^2-s^2}}f(r)\;,
\label{eq:13}
\end{equation} 
which will be more convenient for our purposes. We rewrite the last equation
\begin{equation}
  \frac{2}{\pi}g(s) = \int_0^\infty\mathrm{d} r\,K(s|r)f(r)\;,\quad
  K(s|r) = \frac{2}{\pi} \frac{r}{\sqrt{r^{2}-s^2}}\Theta(r^{2}-s^2)\;,
\label{eq:14}
\end{equation} 
where the factor $2/\pi$ was introduced to ensure that the kernel $K$ is normalised,
\begin{equation}
  \int_0^\infty\mathrm{d} s\,K(s|r) = 1\;,
\label{eq:15}
\end{equation}
with respect to integration over $s$.

Richardson-Lucy deprojection begins with the generalised convolution relation
\begin{equation}
  g(y) = \int\mathrm{d} x\,K(y|x)f(x)\;,
\label{eq:16}
\end{equation}
where the integral kernel $K$ relates the variables $x$ and $y$. By Bayes' theorem, the inverse problem
\begin{equation}
  f(x) = \int\mathrm{d} y\,K'(x|y)g(y)
\label{eq:17}
\end{equation}
has the deconvolution kernel
\begin{equation}
  K'(x|y) = \frac{f(x)}{g(y)}K(y|x)\;,
\label{eq:18}
\end{equation}
provided the deprojection kernel $K$ is normalised as in Eq.~(\ref{eq:14}) and the functions $f$ and $g$ are normalised with respect to the integrals over their domains. \footnote{ \begin{equation*}
 \int {f(x) \mathrm{d} x} = 1 , \ \int {g(y) \mathrm{d} y} = 1                                                                                                                                                                   
\end{equation*}
}  These normalisations are a necessary condition for the algorithm to converge.

Since $f(x)$ is unknown, so is the deconvolution kernel. However, given an estimate $\tilde f_i(x)$ for the function $f(x)$, a corresponding estimate $\tilde g_i(y)$ of the projection is
\begin{equation}
  \tilde g_i(y) = \int\mathrm{d} x\,K(y|x)\tilde f_i(x)\;,
\label{eq:19}
\end{equation} 
allowing us to estimate the deprojection kernel $\tilde K'$ by
\begin{equation}
  \tilde K'(x|y) = \frac{\tilde f_i(x)}{\tilde g_i(y)}K(y|x)\;.
\label{eq:20}
\end{equation} 
Inserting this expression into Eq.~(\ref{eq:17}) gives the updated estimate
\begin{equation}
\label{eq:21}
  \tilde f_{i+1}(x) = \tilde f_i(x)\int\mathrm{d} y\,\frac{g(y)}{\tilde g_i(y)}K(y|x)
\end{equation} 
for the function $f(x)$ given the convolved function $g(y)$, which will later represent the data. Beginning with a reasonable guess for $\tilde f_0(x)$, the iteration Eq. (\ref{eq:21}) usually quickly converges.

A regularisation term needs to be included in presence of noise to prevent overfitting. \citet{Lucy1994} showed that, provided $g(y)$ is normalised, the deprojection algorithm described by the iteration (\ref{eq:21}) can be cast into the form
\begin{eqnarray}
  \Delta_H\tilde f_i(x) &=& \tilde f_{i+1}(x)-\tilde f_i(x) \nonumber\\
  &=& \tilde f_i(x)\left[
    \frac{\delta H[\tilde f_i]}{\delta\tilde f_i(x)}-
    \int\mathrm{d} x\,\tilde f_i(x)\frac{\delta H[\tilde f_i]}{\delta\tilde f_i(x)}
  \right]
\label{eq:22}
\end{eqnarray}
containing the functional derivative of
\begin{equation}
  H[\tilde f] = \int\mathrm{d} y\,g(y)\ln\tilde g(y)
\label{eq:23}
\end{equation}
with respect to $\tilde f_i(x)$. Where $H[\tilde f]$ is equivalent to a likelihood function, which is maximised in order to obtain the best possible solution. He suggested to augment $H[\tilde f]$ by the entropic term
\begin{equation}
  S[\tilde f] = -\int\mathrm{d} x\,\tilde f(x)\ln\frac{\tilde f(x)}{\chi(x)}
\label{eq:24}
\end{equation}
containing a prior $\chi(x)$, to suppress small scale fluctuations. The functional $H$ is then replaced by
\begin{equation}
  H[\tilde f] \to Q[\tilde f] = H[\tilde f] + \alpha S[\tilde f]
\label{eq:25}
\end{equation}
with a parameter $\alpha$ controlling the influence of the entropic term. Since
\begin{equation}
  \frac{\delta S[\tilde f_i]}{\delta\tilde f_i(x)} = -\ln\frac{\tilde f_i}{\chi}-1\;,
\label{eq:26}
\end{equation}
the entropic term changes the iteration prescription to
\begin{equation}
  \Delta\tilde f_i = \Delta_H\tilde f_i+\Delta_S\tilde f_i
\label{eq:27}
\end{equation}
with
\begin{equation}
  \Delta_S\tilde f_i = -\alpha\tilde f_i\left[\ln\frac{\tilde f_i}{\chi}+S\right]\;,
\label{eq:28}
\end{equation}
provided $\tilde f_i$ is also normalised.

This procedure completes our algorithm: We specialise the general deprojection kernel $K(y|x)$ to the projection kernel $K(s|r)$ defined in Eq.~(\ref{eq:14}), identify $x$ with the radius $r$ and $y$ with the projected radius $s$. The function $g(s)$ is replaced by the measured X-ray surface brightness profile $S_{X}(s)$. Then, Eqs.~(\ref{eq:22}), (\ref{eq:27}) and (\ref{eq:28}) allow us to iteratively reconstruct an estimate $\tilde j_{X}(r)$ for the three-dimensional emissivity profile $j_{X}(r)$, including an entropic regularisation term comparing the estimate $\tilde j_{X}(r)$ to a prior $\chi(r)$. Recall that $S_{X}(s)$ and $\tilde j_{X}(r)$ are assumed to be normalised as well as the kernel $K(s|r)$. The complete iteration including the entropic regularisation term reads
\begin{equation}
  \Delta\tilde j_{X,i} = \tilde j_{X,i}\left[
    \int\mathrm{d} s\,\frac{S_X(s)}{\tilde S_{X,i}(s)}K(s|r)-1
    -\alpha\left(\ln\frac{\tilde j_{X,i}}{\chi}+S\right)
  \right]\;,
\label{eq:29}
\end{equation}
with
\begin{equation}
  \tilde S_{X,i}(s) = \int\mathrm{d} r\,K(s|r)\tilde j_{X,i}(r)\;,\quad
  K(s|r) = \frac{2r}{\pi}\frac{\Theta(r^{2}-s^2)}{\sqrt{r^{2}-s^2}}
\label{eq:30}
\end{equation}
and
\begin{equation}
  S[\tilde j_{X,i}] = -\int\mathrm{d} r\,\tilde j_{X,i}(r)\ln\frac{\tilde j_{X,i}}{\chi_i}\;.
\label{eq:31}
\end{equation}

The deprojection begins with a first guess $\tilde j_{X,0}(r)$ for the X-ray emissivity profile, the prior $\chi(r)$ against which the deprojection is to be regularised, and a parameter $\alpha$ controlling the degree of regularisation.	
	
The choice of a constant prior, $\chi$, leads to a statistical bias in the estimates of the deprojected functions such that they appear flatter than they should \citep{Narayan1986, Lahav1989, Lucy1994}.

This issue can be addressed by selecting as a default solution a smoothed version of the obtained result. This approximation, known as floating default \citep{Horne1985, Lucy1994}, is built by adopting the following definition:
\begin{equation}
   \chi(r) = \int_0^{\infty}{ P(r|r') f(r')dr'},
   \label{eq:32}
\end{equation}

where $P(r|r')$ is a normalised, sharply peaked, symmetric function of $r-r'$ and $f(r')$ corresponds to $\tilde j_{X}(r)$. In our investigations, we decided to choose a (properly normalised) Gaussian form with smoothing scale $L$:

\begin{equation*} 
 P(r|r') \sim \exp \left(- \frac{(r-r')^2}{L^{2}}\right).
\end{equation*}
For practical considerations, we are working with discretised data sets and therefore the above integral formulation has to be approximated by sums.

\section{Numerical tests}
\subsection{Simulating X-ray observations}

\begin{figure}[!ht]
\centering
\includegraphics[width=0.90\linewidth]{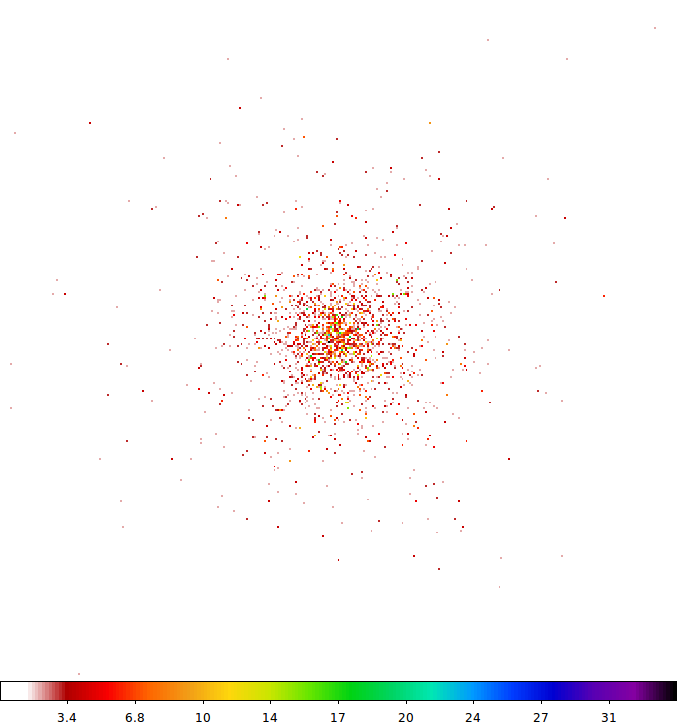} 
\caption{Simulated image of photon counts from a galaxy cluster with a mass of $5\times 10^{14}\;h^{-1}M_{\odot}$, a redshift of $0.2$ and an exposure time of $1000\,$s. The detected photons have energies in the range of $0.5 - 8$ keV. The resolution of this image was lowered by a factor of $6$ compared to the resolution of the simulated Chandra image for visibility.}
\label{fig:cluster_image}
\end{figure}

The goal of this paper is to show that the algorithm defined in the previous section allows to recover the projected gravitational potential of galaxy clusters from their X-ray surface brightness profile, assuming spherical symmetry and hydrostatic equilibrium. For testing this method we simulate galaxy clusters for which we choose a flat standard $\Lambda$CDM model with $\Omega_m = 0.3$, $\Omega_b = 0.04$ and $\Omega_{\Lambda} = 0.7$. For the dark matter, building up the cluster potential well, we use an NFW \citep{NFW1997} density profile
\begin{equation}
\rho(r) = \frac{\rho_{s}}{(r/r_{s}) (1+r/r_{s})^{2}},
   \label{eq:33}
\end{equation}
with the scale radius $r_{s}$ and the characteristic density of the halo. We choose the concentration parameter $c = 5.0$. The gas-mass fraction is set equal to the universal baryon mass fraction $f_b = \Omega_b/\Omega_m$. The gas is assumed to consist of 75\% hydrogen and 25\% helium, both completely ionised, with an effective adiabatic index of $\gamma = 1.2$. 

We approximated the virial radius $r_{\mathrm{vir}}$ of the simulated clusters with $r_{200}$, which is defined to be the radius within which the cluster's mean density is equal to $200$ times the universal critical density at the given redshift,
\begin{equation}
   r_{200} = \left( \frac{3 \ M_{200}}{4 \pi \ 200 \rho_cr(z)} \right)^{1/3}.
   \label{eq:34}
\end{equation}
The gas density and temperature profiles are then calculated, using Eqs.~(\ref{eq:06}) and (\ref{eq:08}). To obtain a temperature profile which drops to zero at a large radius, we choose a large cut-off radius for the gravitational potential of $r_{\mathrm{cut}} = 100\cdot r_{200}$.
The frequency-dependent emissivity due to bremsstrahlung is given by 
\begin{equation}
    j_X(\nu) = C \rho^2 T^{-1/2} \mathrm{exp}\left(-\frac{h\nu}{k_BT} \right) .
    \label{eq:35}
\end{equation} 

The expectation value for the number of photons, emitted in a detectable energy interval $\lbrack E_0$, $E_1\rbrack$ per unit volume and time then reads
\begin{equation}
   N_X = \int_{E_0\left(1+\mathbf{z_{\mathrm{cl}}}\right)}^{E_1\left(1+\mathbf{z_{\mathrm{cl}}}\right)} \frac{\mathrm{d}\left(h\nu\right)}{h} \frac{j_X(\nu)}{h\nu}\;,
   \label{eq:36}
\end{equation}
with the cluster's redshift $z_{\mathrm{cl}}$.

In order to obtain an image comparable to observations we simulate the CCD as follows:
\begin{itemize}
\item We neglect the convolution of the image with the telescope beam: Each pixel is mapped to a unique solid-angle element. The physical area $\delta A$ imaged by one pixel is
\begin{equation}
  \delta A = \delta \theta^2 D_{\mathrm{ang}}^2,
  \label{eq:37}
\end{equation}
where $D_{\mathrm{ang}}$ is the angular diameter distance of the cluster and $\delta \theta$ the angular side length of one pixel, assuming them to be perfectly quadratic.
\item We did not include any cooling effects on the intracluster plasma even though cooling may steepen the X-ray surface brightness profile near the cluster core.
\item Any absorption of X-ray photons between the cluster and the telescope is neglected.
\item The detector has a perfect quantum efficiency within a sharp energy interval. Given the photon counts of the cluster given by Eq.~(\ref{eq:36}), a pixel centred on the radial coordinate $s$ is expected to collect 
\begin{equation}
  \delta N \left(s\right) = \delta A \ \int \mathrm{d}z N_X\left[\sqrt{s^2+z^2}\right] \frac{A_{\mathrm{eff}}}{4 \pi D_{\mathrm{lum}}^2(z_{\mathrm{cl}})} \left(1+z_{\mathrm{cl}}\right)
\label{eq:38}
\end{equation}
photons per second. $D_{\mathrm{lum}}$ is the luminosity distance to the cluster and $A_{\mathrm{eff}}$ is the effective telescope or detector area. Since Eq. (\ref{eq:38}) inherits a conversion from photon energy to photon counts, only one factor of $(1+z_{\mathrm{cl}})$ appears.
\item The limited energy resolution of the telescope is mimicked by choosing appropriate energy intervals in Eq.~(\ref{eq:36}).
\end{itemize}
For the exact properties of the CCD, we adopt the same characteristics as the Chandra Advanced CCD Imaging Spectrometer (ACIS). The detection energy range is set to $0.5 - 8$\,keV. We choose $\left(E_{\mathrm{up}}-E_{\mathrm{low}}\right) /  E_{\mathrm{res}} = 15$ energy intervals, since our method is sensitive only to total numbers of photons and no spectral information is needed.
We calculate photon numbers to pixels by drawing Poisson deviates with the appropriate expectation value $\delta N$ for all 15 energy intervals.
The mean energy $(E_{1} - E_{0})/2$ is assigned to each photon and the sum of energies allotted to the corresponding pixel.
We include statistical noise by adding a constant background such that approximately $15\,\%$ of the detected photons is due to background.

The pixel width is taken to be 0.5 arcseconds and the exposure time is set to $1000\,$s.
In this way we obtain an X-ray surface brightness which is then azimuthally averaged around the centre of the cluster and binned. This profile is used as an estimate for the X-ray surface brightness profile and supplied to the Richardson-Lucy deprojection algorithm described above.

\begin{figure}[!ht]
\centering
\includegraphics[width=1.0\linewidth]{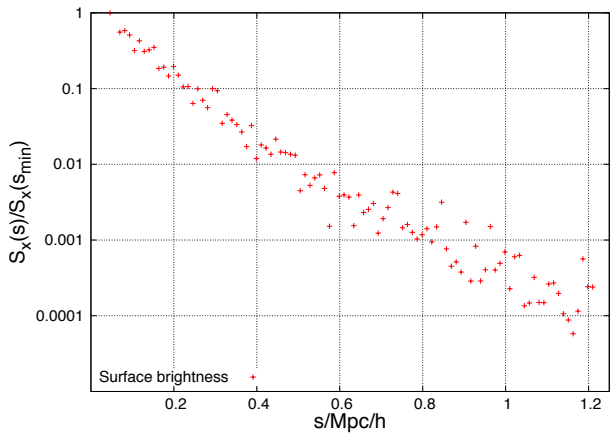} 
\caption{Azimuthally averaged and normalised surface brightness profile of a simulated galaxy cluster with a mass of $5\times 10^{14}\;h^{-1}M_{\odot}$ and a redshift of $0.2$. The corresponding synthetic observational data is shown in Fig.~\ref{fig:cluster_image}.}
\label{fig:cluster_profile}
\end{figure}

Figures~\ref{fig:cluster_image} and \ref{fig:cluster_profile} show the photon counts and the normalised surface brightness of the synthetic observation and the normalised surface brightness profile of a simulated Chandra image for one realization of a galaxy cluster with a mass of $5\times 10^{14}\;h^{-1}M_{\odot}$ and a redshift of $0.2$. With these characteristics and the cosmology given above, this cluster has a scale radius of approximately $r_\mathrm{s} = 0.25$~$h^{-1}\mathrm{Mpc}$ and a virial radius of $r_{\mathrm{vir}} = 1.2$~$h^{-1}\mathrm{Mpc}$. 

% \FloatBarrier

\begin{figure*}[!ht]
\centering
\subfigure[]{\includegraphics[width=0.48\linewidth]{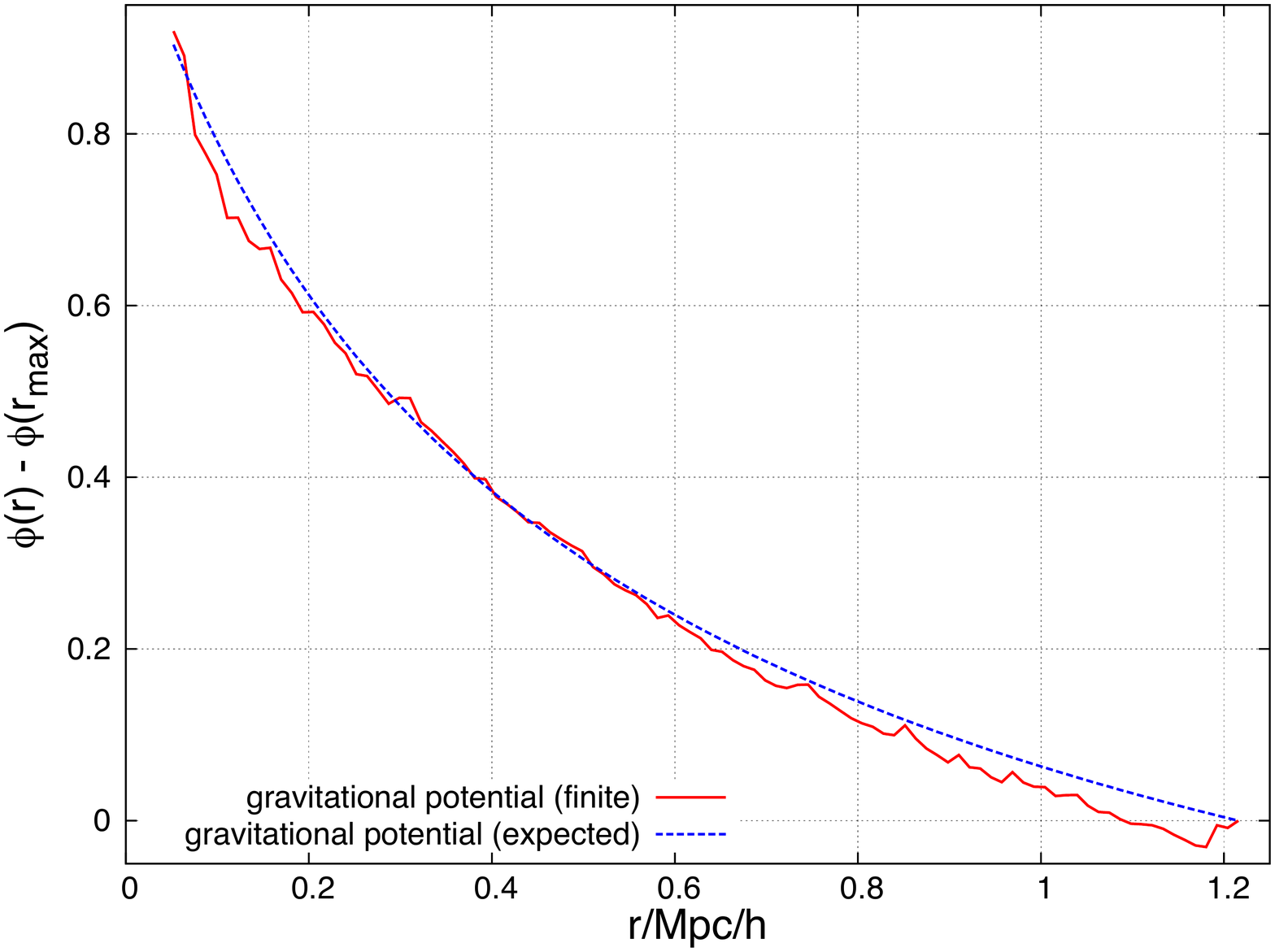}}\qquad \subfigure[]{\includegraphics[width=0.48\linewidth]{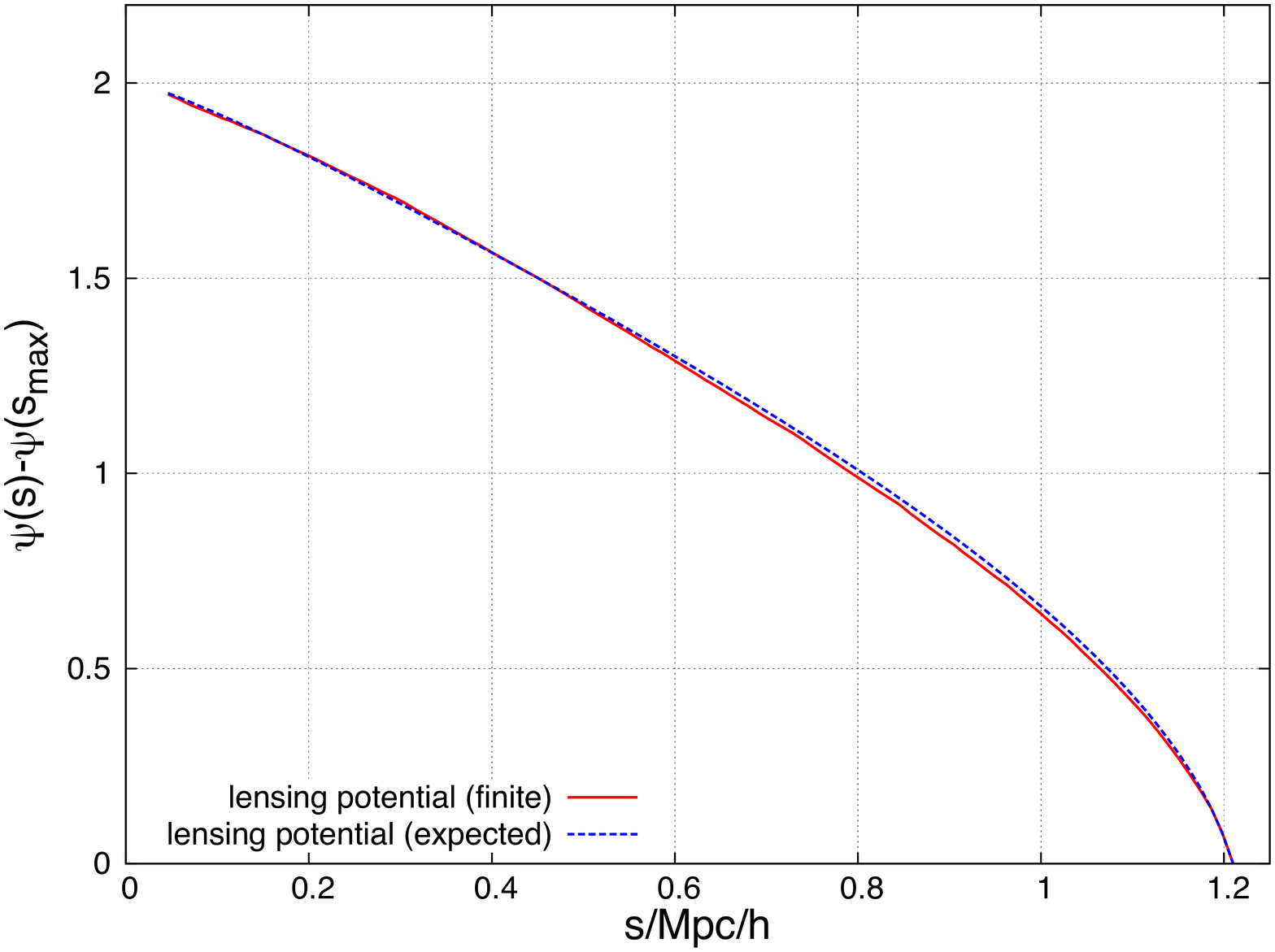}}
\caption{(a) Reconstructed and normalised gravitational potential of a simulated galaxy cluster with a mass of $5\times 10^{14}\; h^{-1}M_{\odot}$ and a redshift of $0.2$. The potential was reconstructed assuming $\alpha = 0.4$ and $L = 0.3\;h^{-1}\mathrm{Mpc}$. (b) Corresponding lensing potential for this simulated cluster.}
\label{fig:lensingpotential}
\end{figure*}

The reconstructed and normalised gravitational potential $\phi$ is shown in Fig.~\ref{fig:lensingpotential}a together with the true potential, the clusters were modeled with. Despite the statistical fluctuations of the surface brightness profile supplied to the algorithm, the contribution of the background noise exceeds the real surface brightness profile at large radii (i.e.~$s\gtrsim0.8\;h^{-1}\mathrm{Mpc}$) which then leads to an overestimation of the gravitational potential. This effect and the normalisation condition of the deprojection algorithm leads to a slight underestimation at smaller radii (i.e.~$s\lesssim0.3\;h^{-1}\mathrm{Mpc}$). However, as we are only interested in the lensing potential, major fluctuations are averaged out as seen in Fig.~\ref{fig:lensingpotential}b. 

\subsection{Testing the algorithm}

\begin{figure*}[!ht]
\centering
\subfigure[]{\includegraphics[width=0.48\linewidth]{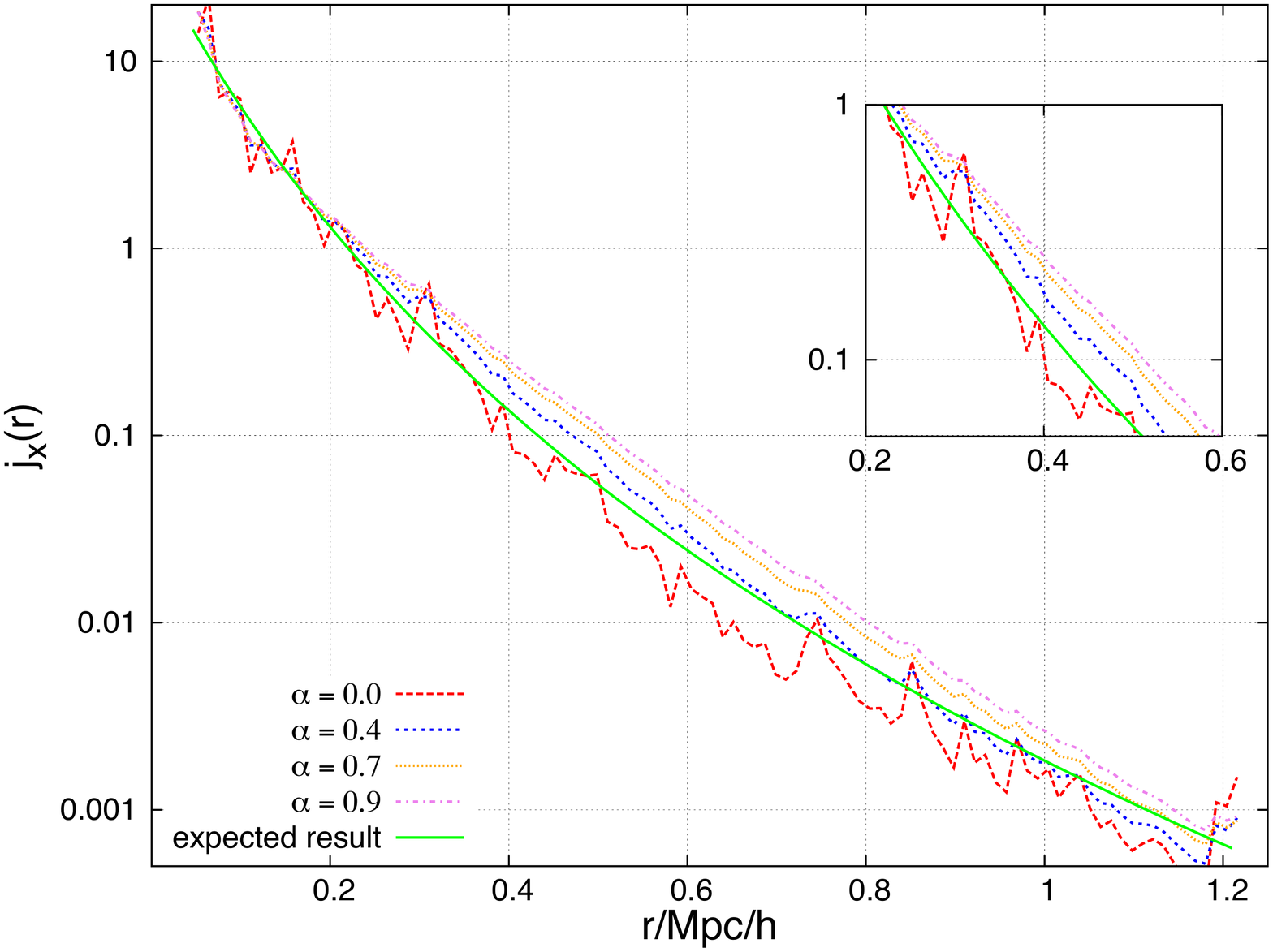}}\qquad \subfigure[]{\includegraphics[width=0.48\linewidth]{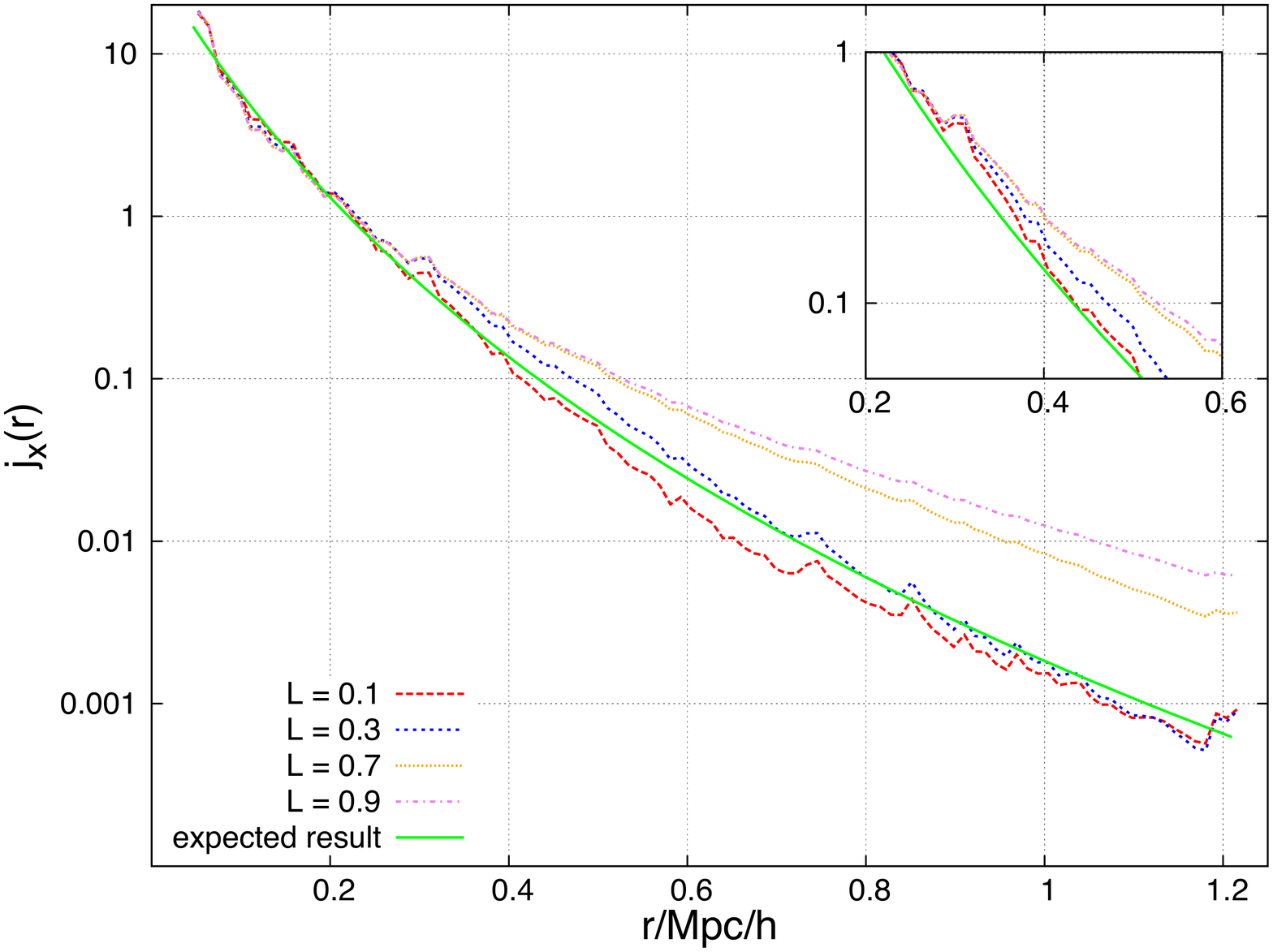}}
\caption{(a) Comparison between different choices for the weights $\alpha$ of the penalty function in the reconstruction of the X-ray emissivity with constant smoothing scale $L=0.3\;h^{-1}\mathrm{Mpc}$. The expected result based on Eq. (\ref{eq:10}) is plotted for reference. Whereas $\alpha = 0.0$ correspond to no regularisation. (b) Comparison between different choices of the smoothing scale of the regularisation function in the reconstruction of the X-ray emissivity and $\alpha = 0.4$. The expected result is plotted for reference.}\label{fig:different_weights_penalty}
\label{fig:alpha_L_compare}
\end{figure*}

Next, we test the sensitivity of the potential reconstruction against changing certain parameters of the deprojection algorithm. 
We compare reconstructions of the X-ray emissivity obtained by assigning different weights $\alpha$ from $[0.0, 0.4, 0.7, 0.9]$ to the entropic regularisation term, with $\alpha = 0.0$ corresponding to no regularisation, and constant $L = 0.3\;h^{-1}\mathrm{Mpc}$.

However, strong regularisation causes the reconstruction to overestimate the signal because of broad averaging by means of the flattening default kernel, because the regularisation penalises curvature. This can be seen at large radii.

A similar conclusion can be drawn from Fig.~\ref{fig:alpha_L_compare}b, which contains the deprojected profile for different choices of the smoothing scale $\mathrm{L}$ from $[0.1, 0.3, 0.7, 0.9]\;h^{-1}\mathrm{Mpc}$ while $\alpha = 0.4$ is kept fixed. Due to the smoothing process performed by the floating default regularisation, a large fraction of the noise pattern, especially for large radii, is averaged out, which improves the convergence towards the expected result.

\begin{figure}[!ht]
\centering
\includegraphics[width=1.0\linewidth]{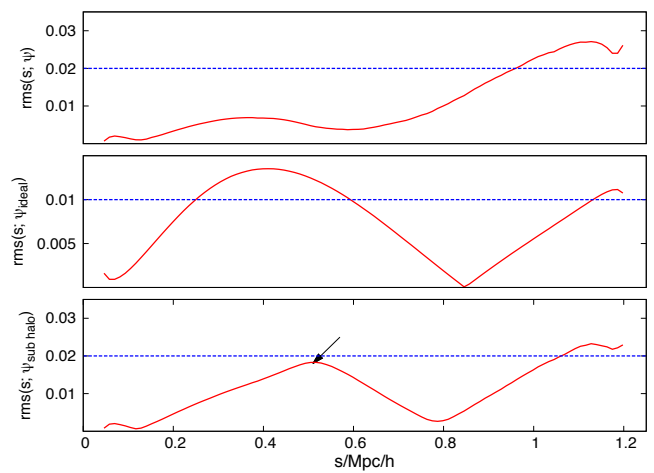} 
\caption{From top to bottom: rms deviation of $\psi$ from $\psi_{\mathrm{true}}$, rms deviation of $\psi_{\mathrm{ideal}}$ from $\psi_{\mathrm{true}}$ and rms deviation of $\psi_{\mathrm{subhalo}}$ from $\psi_{\mathrm{true}}$, according to Eq. (\ref{eq:39}). Calculated from $200$ realisations of a modeled galaxy cluster with a mass of $5\times 10^{14}\;h^{-1}M_{\odot}$, a subhalo with mass $1\times 10^{14}\;h^{-1}M_{\odot}$ respectively, and a redshift of $0.2$. The blue lines represent a $2.0\%$ and $1.0\%$ threshold.}
\label{fig:residua}
\end{figure}

As a final test, we calculate the uncertainty of our reconstructed lensing potentials. We obtain this uncertainty by bootstrapping, sampling the synthetic cluster data $N=200$ times and applying our reconstruction algorithm. For each result of the reconstruction the mean squared deviation from the true potential is calculated and then averaged over the number of bootstraps, giving the rms deviation:
\begin{equation}
	\mathrm{rms}(s; \psi) = \sqrt{\frac{1}{N}\sum_{n=1}^{N} \frac{ \big(\psi^{\mathrm{norm}}(s) - \psi^{\mathrm{norm}}_{\mathrm{true}}(s)\big)^{2}}{{\psi^{\mathrm{norm}}_{\mathrm{true}}(s)}^{2}}}\;,
\label{eq:39}
\end{equation}
where quantities marked with a superscript 'norm' are normalised to reach zero at the maximum projected radius.
These rms are shown in the first panel of Fig.~\ref{fig:residua}. Since this rms incorporates statistical as well as systematic errors, we obtain a relative deviation from the true lensing potential of about $2\,\%$ for large values of $s$. The relative rms increases with increasing radius due to the poorer signal. 

Since we also know the exact surface brightness of the cluster simulation, we can estimate the relative systematic error of the algorithm itself. Doing so, we bin the real surface brightness profile in the same way as we binned our statistical CCD images and obtain a reconstruction that does not inherit any statistical fluctuations or noise. We call this the best possible reconstruction $\psi_{\mathrm{ideal}}$. Its rms with respect to the true lensing potential $\psi_{\mathrm{true}}$ provides an estimate for the systematic error of our reconstruction algorithm (shown in the lower panel of Fig.~\ref{fig:residua}). This mean squared deviation increases slightly with the cluster radius, but always remains below $1.0\,\%$. 

\subsection{Deviations from spherical symmetry}
As a last test we show how deviations from spherical symmetry affect our reconstruction algorithm. 
For this purpose we create a CCD image of a single subhalo with a mass of $1 \times 10^{14}\;h^{-1}M_{\odot}$ at redshift $z_\mathrm{cl} = 0.2$, having a virial radius of $r_\mathrm{subhalo} \approx 0.7\;h^{-1}\mathrm{Mpc}$. We combine this CCD image with the simulated image of the cluster shown previously, with a projected distance of $0.5\;h^{-1}\mathrm{Mpc}$ between the two cluster centres. In a first order approximation we accumulate the surface brightness of these two cluster realisations, before we apply our radial averaging scheme, again assuming spherical symmetry.
Even though the subhalo is clearly visible in the lower panel of Fig.~\ref{fig:residua} (black arrow), we find that the maximum rms deviation of the lensing potential does not exceed the $2.0\%$ deviation achieved without a subhalo.
However, the reconstruction is rather insensitive to this kind of perturbation because of the radial averaging being applied to the input data. Further simulations show that the closer the subhalo lies to the center of
the hosting galaxy cluster the less is the influence on the reconstruction, since the degree of asymmetry decreases.

\section{Conclusions}

This work is motivated by existing and upcoming observational data on galaxy clusters. A set of accurate tools to constrain fundamental cluster quantities is available: weak and strong gravitational lensing observations constrain the line-of-sight projection of the cluster potential, X-ray observations constrain the density and the temperature of the intracluster gas, while the Sunyaev-Zel'dovich effect constrains the gas pressure. A further constraint on the gradient of the gravitational potential can be found by means of galaxy kinematics. Our final goal is to combine all non-lensing information on clusters with existing lensing reconstruction methods \citep[e.g.][]{Bartelmann1996, Bradac2005, Bradac2006, Cacciato2006, Merten2009}.

In this first paper of an intended series, we have outlined a non-parametric reconstruction method for the projected gravitational potential of galaxy clusters from thermal X-ray emission, assuming hydrostatic equilibrium and spherical symmetry. It is more cumbersome, but quite straightforward to extend this method to triaxial halos. This extension is now being addressed.

Our algorithm was laid out in Sect.~2. Assuming that the cluster is near or in hydrostatic equilibrium, we derived an analytic relation between the frequency-integrated bremsstrahlung emissivity and the three-dimensional Newtonian potential, Eq.~(\ref{eq:10}). The algorithm deprojects the observed X-ray surface-brightness profile by means of the Richardson-Lucy method, converts it to the potential and projects that.

Numerical tests show how this algorithm performs under reasonably realistic conditions and how sensitive it is to its parameters. We used one representative cluster realisation to obtain a realistic simulation of observational data. A comparison of different values of the weight $\alpha$ and the smoothing scale $L$ of the regularisation term suggested that there is no significant variation in the final reconstruction of the X-ray emissivity if the floating default regularisation function is adopted with reasonable values.

Even though our simulated galaxy clusters have a rather smooth surface brightness profile, this technique can be applied without restrictions to less well-behaved observational data, e.g.~strongly peaked emission in the cluster centre due to cooling effects. In such cases, the peaked centre could be masked and then passed to the Richardson-Lucy algorithm. The results would still be reliable due to the local character of the reconstruction scheme.

Furthermore, we determined the systematic error of our algorithm by applying it to ideally smooth rather than discretely sampled data to be at most $1.0\,\%$. We finally estimated the combination of systematic and statistical errors of our cluster-reconstruction algorithm to be at most $2.0\,\%$ for $s \approx r_{\mathrm{vir}}$. Both rms were shown in Fig.~\ref{fig:residua}.

We also addressed the problem of a galaxy cluster with an asymmetric surface brightness, e.g. containing a subhalo, in Sect.~3.3. Due to azimuthal averaging, our reconstruction algorithm turns out to be rather insensitive to such kinds of perturbations.
\acknowledgements{This work was supported in part by contract research `Internationale Spitzenforschung II-1' of the Baden-W\"urttemberg Stiftung, by the Collaborative Research Centre TR~33 and project BA 1369/17 of the Deutsche Forschungsgemeinschaft.}

\bibliographystyle{aa}
\bibliography{common_draft}

\end{document}